\documentclass[preprint,12pt]{elsarticle}

\usepackage{amssymb}
\usepackage{amsmath}
\usepackage{algorithmic}
\usepackage{algorithm}
\usepackage{booktabs}
\usepackage{stfloats}
\usepackage{hyperref} 
\usepackage{subfig}
\hypersetup{
hidelinks,
linkcolor=black
}

\begin{document}

\begin{frontmatter}


\title{Line Graph Neural Networks for Link Weight Prediction}
\author[label1]{Jinbi Liang}
\author[label1]{Cunlai Pu\corref{cor1}}
\ead{pucunlai@njust.edu.cn}
\author[label1]{Xiangbo Shu}
\author[label2]{Yongxiang Xia}
\author[label3]{Chengyi Xia}
\cortext[cor1]{Corresponding author.}
\affiliation[label1]{organization={School of Computer Science and Engineering},
            addressline={Nanjing University of Science and Technology},
            city={Nanjing},
            postcode={210094},
            state={Jiangsu},
            country={China}}
\affiliation[label2]{organization={School of Communication Engineering},
            addressline={Hangzhou Dianzi University},
            city={Hangzhou},
            postcode={310018},
            state={Zhejiang},
            country={China}}
\affiliation[label3]{organization={School of Artificial Intelligence},
            addressline={Tiangong University},
            city={Tianjin},
            postcode={300387},
            country={China}}

\begin{abstract}
In real-world networks, predicting the weight (strength) of links is as crucial as predicting the existence of the links themselves. Previous studies have primarily used shallow graph features for link weight prediction, limiting the prediction performance. In this paper, we propose a new link weight prediction method, namely Line Graph Neural Networks for Link Weight Prediction (LGLWP), which learns deeper graph features through deep learning. In our method, we first extract the enclosing subgraph around a target link and then employ a weighted graph labeling algorithm to label the subgraph nodes. Next, we transform the subgraph into the line graph and apply graph convolutional neural networks to learn the node embeddings in the line graph, which can represent the links in the original subgraph. Finally, the node embeddings are fed into a fully-connected neural network to predict the weight of the target link, treated as a regression problem. Our method directly learns link features, surpassing previous methods that splice node features for link weight prediction. Experimental results on six network datasets of various sizes and types demonstrate that our method outperforms state-of-the-art methods.
\end{abstract}



\begin{keyword}
Complex network \sep Link weight prediction \sep Line graph \sep Graph neural network.
\end{keyword}

\end{frontmatter}

\section{Introduction}
Many natural and man-made systems, such as social networks, computer networks, transportation networks, and biological networks, are often investigated using the framework of complex networks~\cite{posfai2016network,newman2018networks}. In this framework, entities are represented by nodes, and the connections between entities are depicted as links. Another critical property of complex networks beyond nodes and links is link weight, which is often underestimated. The weights of links usually signify the strength of connections between nodes. For instance, in a social network, link weights can reflect the level of closeness between friends. In an aviation network, they can represent the volume of flights between cities. In computer networks, link weights can be interpreted as the bandwidth of physical links between routers. Furthermore, link weight has been demonstrated to be one of the factors influencing the evolution of links, communities, and even the overall network topologies~\cite{kossinets2006empirical}.

Nevertheless, in real situations, especially for large networks, it's possible that the weights of some links are missing due to accidents or covered by security considerations. Therefore, an interesting task is to predict the missing link weights to enhance our understanding of complex networks and facilitate various network-related applications. The task of link weight prediction is relatively new compared to link prediction~\cite{qin2023temporal}. The distinction between these two link-related tasks lies in the fact that link weight prediction entails a regression task, whereas link prediction is essentially a classification task.

Previous studies  aiming to tackle  link weight prediction primarily rely on shallow network topological features to estimate connection weights~\cite{lu2010link,zhao2015prediction}. These shallow topological features are typically derived from basic statistical characteristics of the local graph structure and fail to capture more intricate structural and non-structural information. This limitation underscores the need to employ more advanced techniques to develop more effective methods for link weight prediction.

The task of link weight prediction is closely related to link prediction, and as a result, the abundant research on link prediction can offer significant insights into the design of link weight prediction methods. On the one hand, various node similarity metrics have been proposed in link prediction~\cite{lu2011link}, with the assumption that if two nodes share similarities, they have a probability of being connected by a link. Generally, higher values of similarity metrics indicate a greater likelihood of link existence. In fact, these node similarity metrics can be used to generate link features, forming the basic link weight prediction methods~\cite{zhao2015prediction}. On the other hand, to better capture the intricate structural features of networks, graph embedding methods have been introduced in link prediction~\cite{du2019joint,rossi2021knowledge}. These methods transform networks from high-dimensional sparse representations into low-dimensional feature vectors while preserving the structural attributes of nodes and links. Notable examples of graph embedding methods include DeepWalk~\cite{perozzi2014deepwalk} and Node2vec~\cite{grover2016node2vec}. Furthermore, some deep learning-based methods, particularly for graph applications~\cite{wu2020comprehensive,waikhom2023survey}, have been proposed, which can extract task-aware latent graph representation features more accurately. The graph embedding and machine learning techniques originally used in link prediction also hold great potential for application in link weight prediction.
 
\begin{figure}[htbp]
  \centering
  \includegraphics[width=1\linewidth]{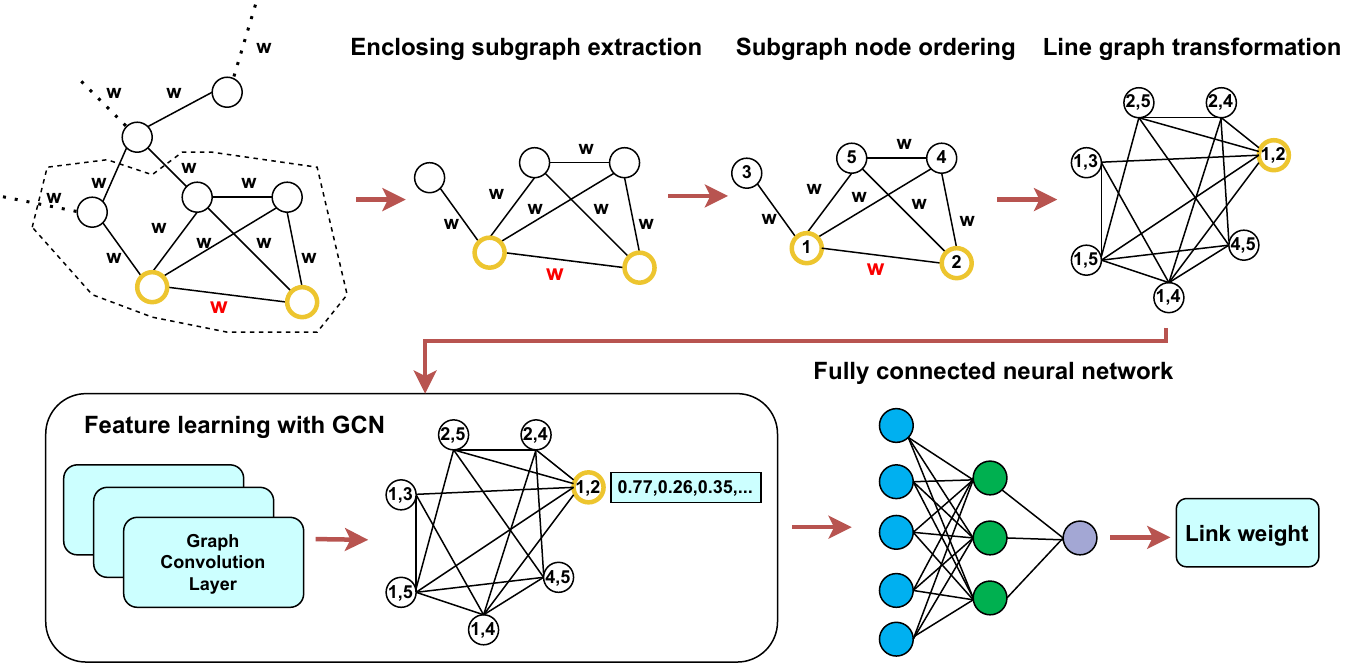}
  \caption{A diagram illustrating the framework of our link weight prediction method, namely, LGLWP. The main components of the method include subgraph extraction, subgraph node ordering, line graph transformation, and feature learning with GCN. }
  \label{fig:1}
\end{figure}

In this paper, we study link weight prediction by leveraging line graph theory and graph neural networks. We employ the line graph theory to transform the task of link feature learning into node feature learning. Utilizing a Graph Convolutional Neural Network (GCN)~\cite{chen2020simple}, we automatically learn the line graph features, which are subsequently employed in the regression task for link weight prediction. The main contributions of our work are summarized as follows:
\begin{enumerate}
	\item We propose a new method for link weight prediction by leveraging the tools of line graph and GCN. Specifically, in this approach, we first extract the subgraph of a target link and label the nodes within the subgraph using an algorithm specifically designed for weighted graph labeling. Subsequently, the labeled subgraph is transformed into the corresponding line graph, which is then fed into the GCN for graph feature learning. Finally, the node features of the line graph are fed into a neural network with two fully connected layers to predict the weight of the target link in the original graph. 
	\item We evaluate the performance of our prediction method across a variety of network datasets, varying in size and type. Comparative experimental results consistently demonstrate that our method achieves state-of-the-art prediction performance. This holds true across different sizes of training sets, highlighting the robustness of our method's performance. 
\end{enumerate}

\section{Related work}
In this section, we provide the related works that are related to our research. We first present the current representative methods for link weight prediction. Subsequently, we introduce graph representation learning and its application in link prediction, which serves as valuable insights for our research. 

\subsection{Link weight prediction}
Link weight prediction is a relatively new research problem within network science.  L\"{u} et al.~\cite{lu2010link} pioneered the research of this problem by introducing the weighted local similarity metrics, such as Weighted Common Neighbors (WCN), Weighted Adamic-Adar (WAA), and Weighted Resource Allocation (WRA). Similarly, Zhao et al.~\cite{zhao2015prediction}  extended the unweighted local similarity metrics to their weighted counterparts using the reliable routing method. These weighted local similarity metrics are known for their simplicity and effectiveness in predicting link weights. Fu et al.~\cite{fu2018link} further incorporated node similarity metrics from the original network and node centrality metrics from the line graph to perform link weight regression prediction. However, it's important to note that these node-similarity-based link-weight prediction methods primarily rely on shallow network structure information, leading to limited prediction performance. Another limitation of these methods is their inability to ensure consistently good performance across different types of networks.  For example, common neighbor-based metrics may work well in social networks or collaboration networks but have been shown to perform poorly in electrical grids and biological networks~\cite{lu2011link}.

Additionally, there are approaches that do not rely on node similarity. For instance, Cao et al.~\cite{cao2020link} introduced a link weight prediction method based on the matrix factoring scheme. In this approach, a link weight matrix is generated by perturbing the structure of an observed weighted network, and another link weight matrix is reconstructed using the factorized latent factors of the observed network. These two matrices are then combined to determine the weights of the target links. Liu et al.~\cite{liu2020new} introduced a novel index called `connection inclination' to capture the actual distribution of link weights. Utilizing this index, they developed a parameterized regression model with an approximately linear time complexity for link weight prediction.

\subsection{Graph representation learning  for link prediction}
 Currently, various graph representation learning methods have been proposed to encode nodes into low-dimensional vectors, capturing both the nodes' positions and their local structures within the graph. Representative methods include DeepWalk~\cite{perozzi2014deepwalk}, node2vec~\cite{grover2016node2vec}, and SDNE~\cite{wang2016structural}. Furthermore, graph deep learning models can automatically derive graph features, with notable examples being Graph Convolutional Networks (GCN)~\cite{kipf2016semi}, Graph Autoencoders (GAE)~\cite{kipf2016variational}, and Variational Graph Autoencoders (VGAE)~\cite{kipf2016variational}.
 
Recently, graph representation learning has been applied to the task of link prediction. For instance, Zhang et al.~\cite{zhang2017weisfeiler} introduced the Weisfeiler-Lehman Neural Machine (WLNM) method, which demonstrated improved performance compared to heuristic methods. This approach involves extracting enclosing subgraphs for potential links, labeling these subgraphs, encoding them as adjacency matrices, and inputting them into a neural network to predict link existence.
Subsequently, the SEAL method~\cite{zhang2018link} was developed for link prediction, enabling the automatic learning of structural features from first-order and second-order enclosing subgraphs. Building upon SEAL, LGLP~\cite{cai2021line} introduced the line graph into the graph neural network model for feature learning, leading to enhanced prediction performance.
In the most recent development, Song et al.~\cite{song2023xgcn} introduced a novel link prediction method called xGCN. This method encodes graph-structured data in an extreme graph-convolutional fashion, making it applicable to large-scale social networks. These advancements underscore the ongoing exploration of sophisticated deep learning and graph embedding techniques in network prediction tasks.

To our knowledge, graph representation learning has not been thoroughly explored in the context of link weight prediction. Inspired by related works on link prediction, we employ graph representation learning in link weight prediction, specifically utilizing GCN to capture deeper graph features for prediction.

\section{Proposed method}
In this section, we shall begin by presenting the problem of link weight prediction. Subsequently, we will introduce our link weight prediction method, named the Line Graph-Based Link Weight Prediction (LGLWP) method. The diagram of our method is provided in Figure \ref{fig:1}, illustrating the four key components: enclosing subgraph extraction, subgraph node ordering, line graph transformation, and feature learning with GCN. Algorithm \ref{alg:1} gives the framework of LGLWP. 

\begin{algorithm}[ht]
    \caption{Link weight prediction procedures of LGLWP.}
    \label{alg:1}
    \textbf{Input}: A target link with two endpoints \(v_1\) and \(v_2\), Graph \(G\) \\
    \textbf{Output}: Predicted link weight
    \begin{algorithmic}[1] 
        \STATE Extract \(\text{1-hop}\) enclosing subgraph \(G_{(v_1,v_2)}\)
        \STATE Label all the nodes in \(G_{(v_1,v_2)}\) with the  weighted graph labeling algorithm \\
        \STATE Generate link attributes using Equation (\ref{eq3})
        \STATE Transform \(G_{(v_1,v_2)}\) to line graph \(L(G_{(v_1,v_2)})\)
        \STATE Learn the node embeddings of \(L(G_{(v_1,v_2)})\) with GCN and predict the link weight with a neural network of two fully connected layers
    \end{algorithmic}
\end{algorithm}

\subsection{Problem description}
Let $G(V,E,W)$ represent an undirected weighted network, where $V$ denotes the node set, $E$ is the link set, and $W$ represents the link weight set consisting of the weight values of all links. In the link weight prediction problem, $W$ is randomly divided into a training set and a test set, denoted as $W_{\text{train}}$ and $W_{\text{test}}$, respectively, where $W_{\text{train}} \cup W_{\text{test}} = W$ and $W_{\text{train}} \cap W_{\text{test}} = \emptyset$. The objective is to effectively predict the link weight values in the test set given the graph $G(V,E,W_{\text{train}})$.
Considering the large variance of the link weight values, it is common to preprocess the link weight data before prediction. Specifically, we use the exponential transformation method to normalize all the link weights to the interval $(0,1)$, i.e.,
\begin{equation}
w^* = e^{-\frac{1}{w}}.
\end{equation}

\subsection{Enclosing subgraph extraction}
The first step in our method is to extract the enclosing subgraphs of all the target links. These subgraphs serve as our reference for link weight prediction. Generally, a larger subgraph can provide more information for prediction, but it also comes with higher computational cost. To strike a balance between prediction performance and computational efficiency, we choose to extract only one-hop subgraphs for all target links. The one-hop subgraph of a target link with two endpoints $v_i$ and $v_j$ is denoted as $G_{v_i,v_j}$. In addition to nodes $v_i$ and $v_j$, this subgraph includes all the neighbors of nodes $v_i$ and $v_j$, as well as all the links between these nodes. 

We typically anticipate a subgraph to contain sufficient information to make accurate predictions while avoiding excessive size that can result in the inclusion of noisy and irrelevant nodes.  Accordingly, we set the number of nodes in each subgraph to be less than $n$. Specifically, if there are more than $n$ nodes in a subgraph, we randomly select $n$ nodes and remove all other nodes and their links. Otherwise, we keep all the nodes and links in the   subgraph. This ensures that the subgraphs of all target links have similar sizes, facilitating the evaluation and comparison of prediction performance.

\subsection{Subgraph node ordering}
The second step of our method is node labeling in the extracted enclosing subgraph. The objective of this step is to establish a consistent  node order that  reflects their relative positions within the subgraph and their structural roles in link weight prediction. To achieve this, we employ the node labeling method proposed by Zulaika et al.~\cite{zulaika2022lwp}. This method is  essentially an adaptation of the 1-dimensional Weisfeiler-Lehman (WL) algorithm~\cite{weisfeilerreduction}, specifically  tailored to handle weighted graphs. 
 The method organizes the node order based on their weighted distances and the weighted distances of their neighbors to the target nodes. This arrangement preserves the topological directionality of nodes to the target link, thereby  facilitating the task of learning and predicting  link weights. 

\begin{figure}[htbp]
  \centering
  \includegraphics[width=0.9\linewidth]{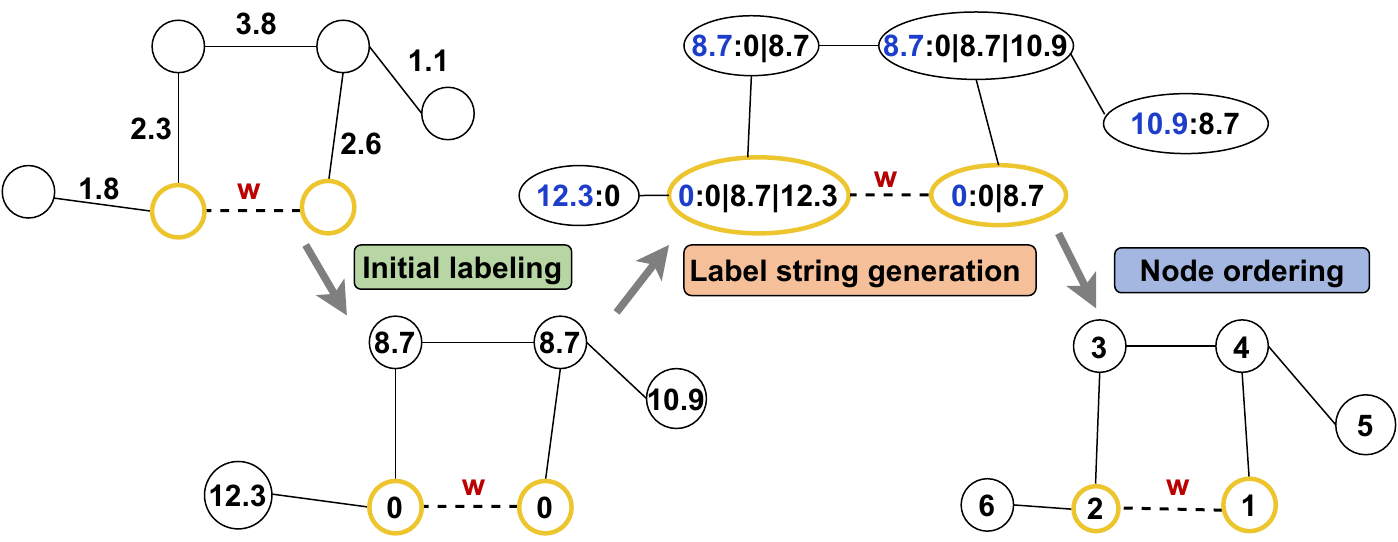}
  \caption{A diagram  illustrating the subgraph node ordering algorithm, where the dashed line is the target link (whose weight is unknown) with two target nodes marked with yellow circles. }
  \label{fig:2}
\end{figure}

 As depicted in Figure \ref{fig:2}, the two nodes of a target link are initially labeled as 0. For a non-target node, there exist two  smallest weight paths  to  each of the two target nodes, and the label of this node is set to be the sum of the link weights along these two paths.    Subsequently, the label of each node is updated using a unique label string, which is a combination of the initial label of the node and the initial labels of its neighboring nodes in increasing order. Ultimately, by comparing the label strings, we obtain the  node ordering, with the target nodes always appearing as the first two in the order.

In the  weighted adjacency matrix of the ordered subgraph, the weight value of the target link  is set to be invisible (technically negative one), and our task is to predict this weight value. Furthermore, the row vectors of this matrix are utilized as the attribute vectors for nodes. For instance, the first row of this matrix provides the weights of the links connected to node 1, making it  the attribute vector of node 1. 

\subsection{Line graph transformation}
\begin{figure}[htbp]
  \centering
  \includegraphics[width=0.9\linewidth]{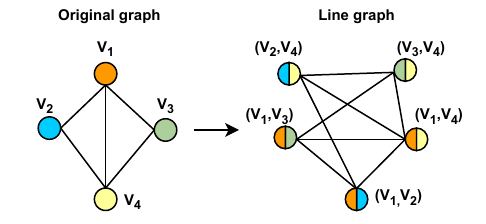}
  \caption{A diagram illustrating the line graph transformation, where the link between nodes $v_i$ and $v_j$ in the original graph corresponds to the node $(v_i, v_j)$ in the line graph. }
  \label{fig:3}
\end{figure}

In graph theory, the line graph $L(G)$ is the dual of the original graph $G$~\cite{gross2018graph}. In  $L(G)$, each node corresponds to a unique link in $G$, and each link in $L(G)$ corresponds to an adjacency of two links in $G$.  An example of a graph and its corresponding line graph is  shown in Figure \ref{fig:3}.

Since the node labels in the ordered subgraph embody structural information contributing to link weight prediction, it is necessary to transfer this information to the corresponding line graph. Therefore, we label a node in the line graph as $(v_i, v_j)$ if it corresponds to the link between nodes $v_i$ and $v_j$ in the original graph.  
Moreover, the node attributes in the original subgraph, which are represented by the row vectors of the  weighted adjacency matrix, contain both topology and weight information for link weight prediction. Thus, we also need to transfer this information   into the line graph.   
To achieve this,  we employ a concatenation operation that converts the node attributes in the original subgraph into the node attributes in the corresponding line graph.    The concatenation operation is defined as
\begin{equation} \label{eq3}
	\resizebox{0.8\linewidth}{!}{$h((v_i,v_j))=\mathrm{concate}(h(\min(v_i,v_j)),h(\max(v_i,v_j))),$}
\end{equation}
where $h(\cdot)$ represents attribute of a component; $v_i$ and $v_j$ are  two connected nodes in original subgraph, and $(v_i,v_j)$ is a node  in the line graph corresponding to the link between nodes $v_i$ and $v_j$ in the original graph; $\mathrm{concate}(\cdot)$ represents a concatenation operation for the two inputs. This concatenation operation ensures that the attribute of a node in the line graph is the concatenation of the attributes of the two endpoints (in increasing order of labels) of the corresponding link in the original graph. 

By transforming the ordered subgraph into the corresponding line graph, link weight prediction in the original graph is transformed into the task of `node weight’ prediction in the line graph. This transformation will not significantly increase computational complexity but will facilitate the usage of machine learning techniques.  

\subsection{Feature Learning with GCN}
Deep learning methods have been successfully applied in many fields such as image processing and speech recognition. However, many traditional deep learning methods do not work well in extracting features from graphs that exist in non-Euclidean space. Kipf et al.~\cite{kipf2016semi} proposed a multilayer graph convolutional neural network that can be directly applied to graph data. This learning method aggregates information from a node and its neighbors, generating new node embeddings that contain rich neighborhood information.

In this work, we use a GCN to learn node embeddings for the line graph, where each node represents a link in the original subgraph. The obtained node embeddings of the line graph are then used to predict the weight of the   target link in the original subgraph.

Let $L\left(G_{v_1,v_2}\right)$ be the line graph representation of the enclosing subgraph $G_{v_1,v_2}$ of the target link with endpoints $v_1$ and $v_2$. Let $\left(v_i,v_j\right)$ be a label of a node in the line graph corresponding to the link in the original subgraph with endpoints $v_i$ and $v_j$. 
Assuming that the node embedding of $\left(v_i,v_j\right)$ in the $k$-th layer of the GCN is denoted as $Z_{\left(v_i,v_j\right)}^{\left(k\right)}$, then the node embedding of $\left(v_i,v_j\right)$ in the $\left(k+1\right)$-th layer~\cite{kipf2016semi} is
 \begin{equation}
	Z_{\left(v_i,v_j\right)}^{(k+1)}=\sigma(\widetilde{D}^{-\frac12}\widetilde{A}\widetilde{D}^{-\frac12}Z_{\left(v_i,v_j\right)}^{(k)}W^{(k)}),
\end{equation}
where $\widetilde{A}=A+I_{N}$, with $A$ being the adjacency matrix of  $L\left(G_{v1,v2}\right)$ and $I_N$ being an $N \times N$ identity matrix; $\widetilde{D}$ is a degree matrix with $\widetilde{D}_{ii}=\sum_j\widetilde{A}_{ij}$; $W^{(k)}$ is the trainable parameter matrix of the $k$-th layer of the GCN; and $\sigma(\cdot)$ is the activation function used in each layer of the GCN.

The input for the first layer of the GCN, i.e., $Z_{(v_i,v_j)}^0=h{((v_i,v_j))}$,  comprises all node attributes of the line graph.  The output of the GCN is the node embeddings of the line graph. Finally, these node embeddings are used to predict the link weights in the original subgraph. Specifically,  this task is treated as a standard regression task performed by a two-layer fully connected neural network,  trained by minimizing the root-mean-square error loss.

\section{Experiments}

\subsection{Datasets description}
 We evaluate our proposed method on six weighted networks of different types and sizes. The topological characteristics of these networks are provided in Table \ref{tab:1}. 
 
\begin{itemize}
 \item Neural-net~\cite{watts1998collective}: A neural network of C. elegans with connections between neurons by either a synapse or a gap junction. The weights of links represent the numbers of interactions between the neurons.
 \item C. elegans~\cite{liu2013correlations}: A network that describes the interactions between metabolites in the C. elegans. In this network, the links represent the connections between pairwise metabolites. The weights assigned to the links reflect the occurrence of multiple interactions between these metabolites. 
\item Netscience~\cite{newman2006finding}: The largest component of a co-authorship network collected by M. Newman, where scientists collaborate on the studies of network science. The weights of links in this network are calculated based on the information of co-authored papers and co-authors. 
\item P. blogs~\cite{adamic2005political}: A network depicting hyperlinks between political web blogs during the 2004 US Election. We remove the directions of the links and aggregate the links if they have the same endpoints. The link weight represents the volume of hyperlinks between blogs. 
\item UC-net~\cite{opsahl2009clustering}: A network that describes communication between users within an online student community at the University of California, Irvine. Users are represented as nodes, while directed links indicate the flow of sent messages.  We remove the directions of the links and aggregate links with the same endpoints. The weight of each link denotes the number of messages.
\item Condmat~\cite{leskovec2007graph}: A network representing collaborations among scientists who  posted preprints on  condensed matter research at www.arxiv.org between 1995 and 1999. The nodes are scientists, and the links represent the co-authorship among them. The weights assigned to the network follow the methodology outlined in the original paper. 
\end{itemize}
\begin{table}
    \resizebox{0.9\linewidth}{!}{
	\begin{tabular}{cccc ccc}
        \toprule
		Datasets & $\left|V\right|$ & $\mid E\mid $ & Range of weights & Categories \\
		\midrule
		Neural-net & 296 & 2,137 &[1, 72] &biology \\
		C. elegans & 453 & 2,025 & [1, 114]&biology \\
		Netscience & 575 & 1,028 &[0.0526, 2.5] &coauthorship \\
		P. blog & 1224 & 16,715 &[1, 3] &social \\
		UC-net & 1899 & 13,838 &[1,184] &social \\
		Condmat & 16264 & 47,594 &[0.058824, 22.3333] &coauthorship \\
		\bottomrule
	\end{tabular}
    }
    \caption{Basic topological features of the weighted networks}
	\label{tab:1}  
\end{table}

\subsection{Evaluation metric}
In the literature of link weight prediction, researchers usually employ Pearson's correlation coefficient and root-mean-square error (RMSE) as evaluation metrics. However, it has been pointed out   that the two metrics have practically equal evaluative power~\cite{liu2023self}. Therefore, we choose only RMSE to measure the prediction performance.
The definition of RMSE is
\begin{equation} RMSE = \sqrt{\frac{\sum_{i=1}^n(y_i - \hat{y}_i)^2}{n}}, \end{equation}
where $y_i$  and $\hat{y}_i$  are respectively the real weight and   the predicted weight of the $i$-th target link.
Obviously, the smaller the RMSE, the closer the predicted values are to the actual values, indicating higher prediction accuracy for a method.

\subsection{Experimental settings}
In the experiments, we randomly select 90\% of the link weights as the training set and the remaining 10\% as the test set.  Considering the randomness of a single experiment, for each prediction method, we conduct 10 independent divisions of training and test sets and calculate the average performance.  The number of nodes in each extracted subgraph is set to be no greater than $n=10$. The machine used for the experiments is a laptop configured with an i9-12900H 2.50 GHz processor, 16GB RAM, and an Nvidia 3070 GPU.

The parameters in our prediction method are  set empirically. Specifically, we utilize three graph convolution layers to compute the node embeddings, with the output feature dimensions of each layer set to 32. Link weight regression prediction is then performed through two additional fully connected layers. The number of training iterations is adjusted for networks of various sizes. In particular, 5 training epochs are used for large networks, such as Condmat, P. blogs, and UC-net, while 15 training epochs are used for the remaining networks.

\subsection{Baselines}
 To evaluate the performance of our method, LGLWP, we compare it with the well-known methods using the given datasets. We first select seven shallow feature-based link weight prediction methods, which include three reliable routing-based methods (rWCN, rWAA, and rWRA)~\cite{zhao2015prediction}, three line graph-based methods (LG-RF, LG-GBDT, and LG-SVM)~\cite{fu2018link}, and NEW~\cite{liu2020new}.

We then select seven representative graph representation models   including Deepwalk~\cite{perozzi2014deepwalk}, Node2vec~\cite{grover2016node2vec}, Grarep~\cite{cao2015grarep}, SDNE~\cite{wang2016structural}, LINE~\cite{tang2015line}, GAE~\cite{kipf2016variational}, and VGAE~\cite{kipf2016variational}. Since these graph learning models are not specifically designed for link weight prediction, we modified them by concatenating the node embeddings as link attributes, and further trained a linear regression model to predict link weights, following the approach outlined in~\cite{liu2023self}.

In addition, we employ the original GCN model to learn node embeddings, and then use the inner products of the node embeddings to approximate the weights of links~\cite{kipf2016variational}. However, it is essential to note that this GCN-based method is  fundamentally different with our proposed method   as we incorporate steps of subgraph abstraction, node ordering, and line graph transformation.
Finally, SEA~\cite{liu2023self}, a self-attention-enhanced graph self-encoder, is also included to the baselines. SEA improves link weight prediction by learning deep graph features.

\subsection{Results and Analysis}
\begin{table}[ht]
    \centering
    \resizebox{0.9\linewidth}{!}{
	\begin{tabular}{ccccccc}
		 \toprule
	Model & Neural-net        & C. elegans     & Netscience    & P. blogs        & UC-net     & Condmat       \\
	\midrule
	rWCN     & 5.78±0.74     & 1.79±0.76     & 0.43±0.05     & 3.02±0.03     & 2.4527±0.0942 & 0.1992±0.0029 \\
	rWAA     & 6.3±0.75      & 2.36±0.84     & 0.42±0.05     & 0.89±0.02     & 0.6617±0.0292 & 0.1816±0.0031 \\
	rWRA     & 6.7±0.76      & 2.93±0.91     & 0.42±0.05     & 1.09±0.01     & 0.5852±0.0067 & 0.1932±0.0027 \\
	LG-RF    & 0.235±0.006   & 0.183±0.003   & 0.213±0.005   & 0.099 ±0.003  & 0.223±0.002   & -             \\
	LG-GBDT  & 0.383±0.004   & 0.276±0.005   & 0.181±0.006   & 0.239±0.003   & 0.369±0.003   & -             \\
	LG-SVM   & 0.236±0.006   & 0.152±0.004   & 0.212±0.004   & 0.171±0.004   & 0.225±0.003   & -             \\
	NEW      & 0.2056±0.0064 & 0.1421±0.0081 & 0.0891±0.0115 & 0.0797±0.0024 & 0.2076±0.0017 & 0.1953±0.0016 \\
	\midrule
	Deepwalk & 0.2211±0.0043 & 0.1421±0.0045 & 0.1214±0.0151 & 0.0816±0.0023 & 0.2124±0.0026 & 0.1943±0.0008 \\
	Node2vec & 0.2153±0.0054 & 0.1413±0.0052 & 0.1199±0.0126 & 0.0817±0.0021 & 0.2088±0.0022 & 0.2032±0.0011 \\
	Grarep   & 0.2254±0.0092 & 0.1424±0.0053 & 0.1484±0.0378 & 0.0798±0.0021 & 0.2098±0.0012 & 0.1945±0.0016 \\
	SDNE     & 0.2060±0.0058 & 0.1380±0.0167 & 0.1386±0.0263 & 0.0771±0.0029 & 0.2056±0.0029 & 0.1808±0.0014 \\
	LINE     & 0.2222±0.0079 & 0.1390±0.0052 & 0.1377±0.0112 & 0.0809±0.0021 & 0.2102±0.0016 & 0.1927±0.0016 \\
	$\text{GAE}$      & 0.2161±0.0082 & 0.1508±0.0058 & 0.4452±0.0052 & 0.1466±0.0142 & 0.2360±0.0041 & 0.4112±0.0017 \\
	$\text{VGAE}$     & 0.2332±0.0089 & 0.1496±0.0054 & 0.4458±0.0052 & 0.1340±0.0008 & 0.2318±0.0043 & 0.4127±0.0017 \\
        GCN     & 0.2216±0.0098 & 0.1583±0.0139 & 0.1232±0.0146 &  0.2720±0.0770 & 0.254±0.0614 &  0.2117±0.0036 \\
        SEA     & \underline{0.2015±0.0052} & \pmb{0.1113±0.0055} & \underline{0.0823±0.0094} & \pmb{0.0754±0.002} & \pmb{0.1976±0.0028} & \underline{0.1694±0.0018} \\
	\midrule
	LGLWP    & \pmb{0.1859±0.0053} & \underline{0.1279±0.0041} & \pmb{0.0624±0.0137} & \underline{0.0759±0.0019}  & \underline{ 0.1994±0.0036} & \pmb{0.1506±0.0022} \\ 
	\bottomrule         
	\end{tabular}
    }
    \caption{The results of  RMSE along with its standard deviations for 10 trials for all prediction methods have been compared across all datasets. The best results have been bolded, and the second-best results are indicated with underlines.}
   \label{tab:2}  
\end{table}

\begin{table}[ht]
	\centering
    \resizebox{0.9\linewidth}{!}{
		\begin{tabular}{ccccccc}
			\toprule
			Method & Neural-net        & C. elegans     & Netscience    & P. blogs        & UC-net     & Condmat       \\
			\midrule
			Random labeling    & 0.2115 ±0.0066     & 0.1466±0.0068     & 0.0762±0.0154     & 0.0833±0.0013     & 0.2096±0.0025 & 0.1745±0.0034 \\
			Weighted graph labeling      & 0.1859±0.0053     & 0.1279±0.0041     & 0.0624±0.0137     & 0.0759±0.0019     & 0.1994±0.0036 & 0.1506±0.0022 \\
			\bottomrule         
	\end{tabular}
    }
    \caption{The results of RMSE and its standard deviations for 10 trials for random labeling and weighted graph labeling, demonstrating the superiority of the latter.}
	\label{tab:3}  
\end{table}

Table \ref{tab:2} presents the RMSE results, where our method, LGLWP, demonstrates much better performance (indicated by smaller RMSE values) compared to all methods except SEA across all datasets. LGLWP shows comparable performance to SEA in terms of RMSE, with a slight advantage on half of the datasets.

It's important to note that the SEA model requires global graph information in its computation, incurring high computational cost, especially for large graphs. In contrast, LGLWP leverages only the first-order neighboring structure around the target link during computation and directly learns link features through the line graph transformation. This design choice results in significantly lower computational cost for LGLWP compared to SEA.

\begin{figure}[ht]
\begin{minipage}{0.32\linewidth}
 	\vspace{3pt}
 	\centerline{\includegraphics[width=\textwidth]{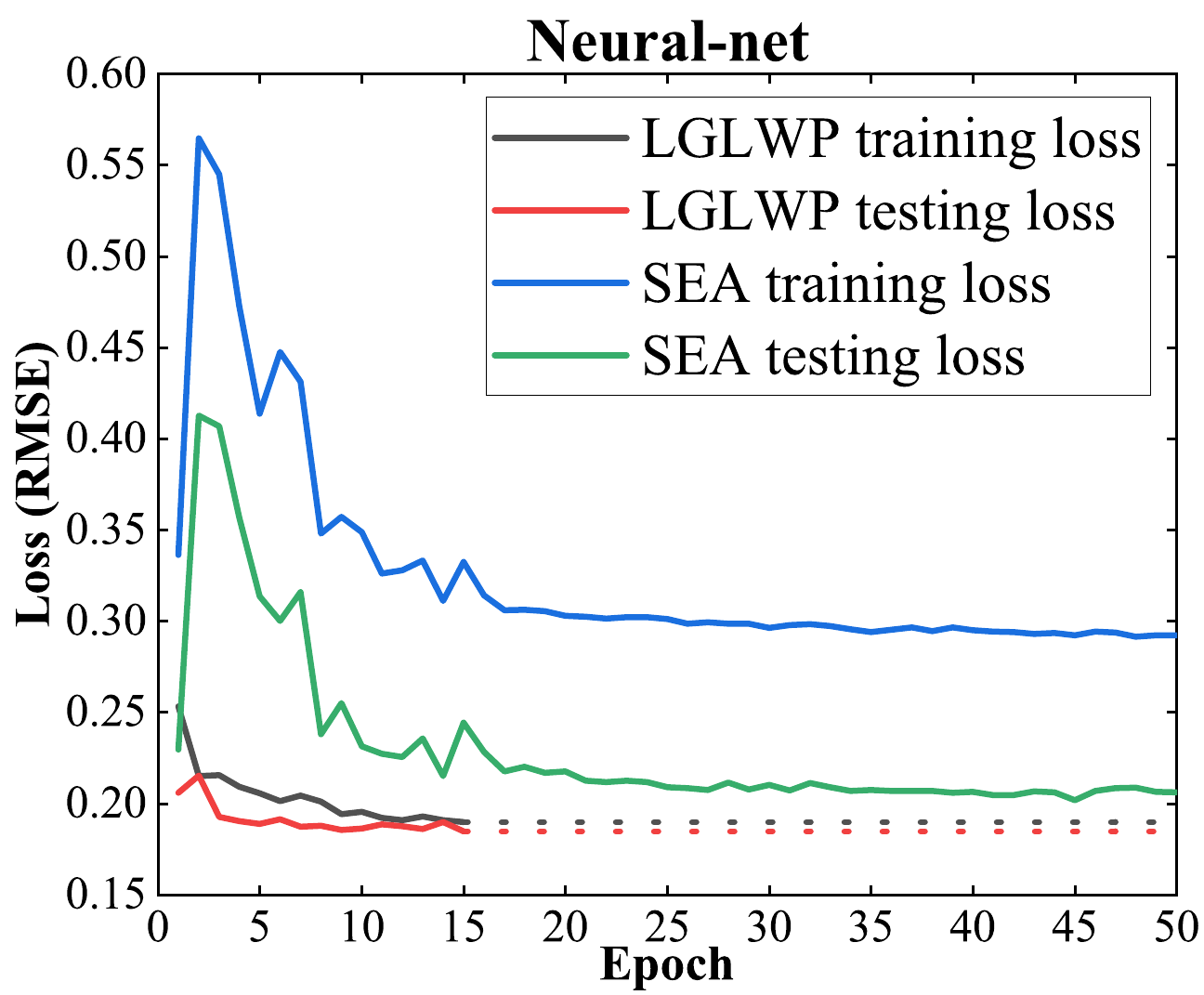}}
 	\vspace{3pt}
 	\centerline{\includegraphics[width=\textwidth]{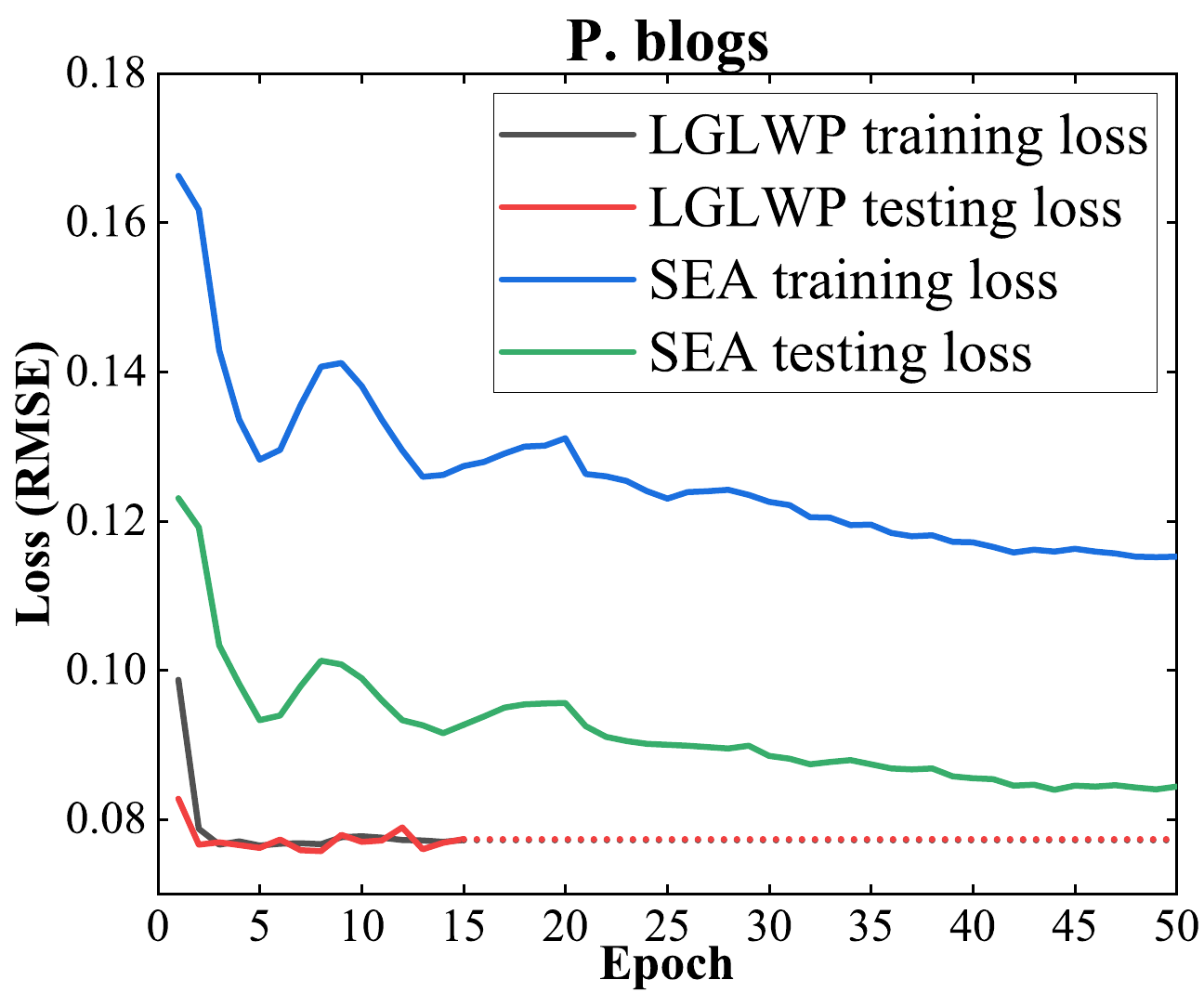}}
 \end{minipage}
\begin{minipage}{0.32\linewidth}
 	\vspace{3pt}
 	\centerline{\includegraphics[width=\textwidth]{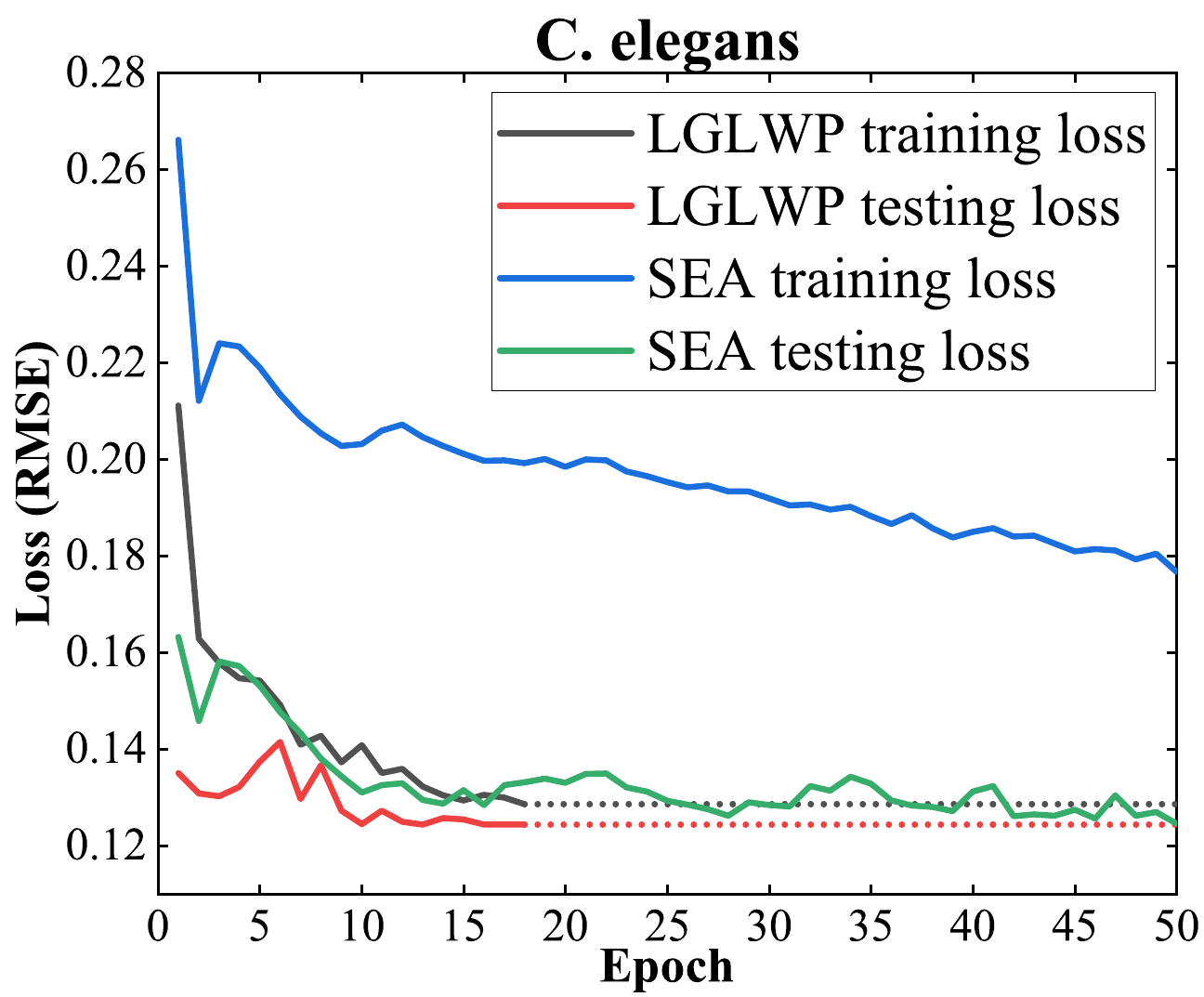}}
 	\vspace{3pt}
 	\centerline{\includegraphics[width=\textwidth]{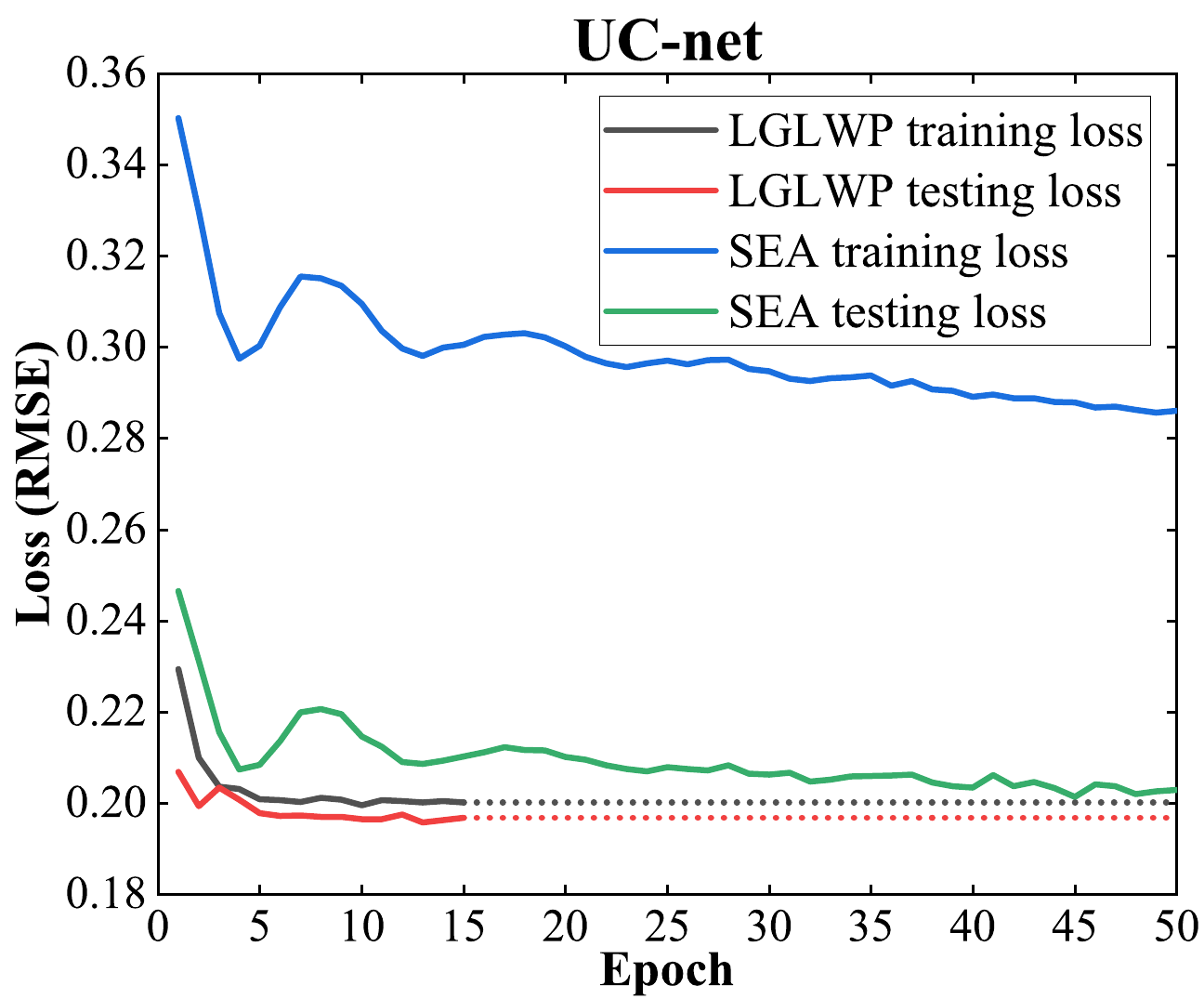}}
 \end{minipage}
 \begin{minipage}{0.32\linewidth}
 	\vspace{3pt}
 	\centerline{\includegraphics[width=\textwidth]{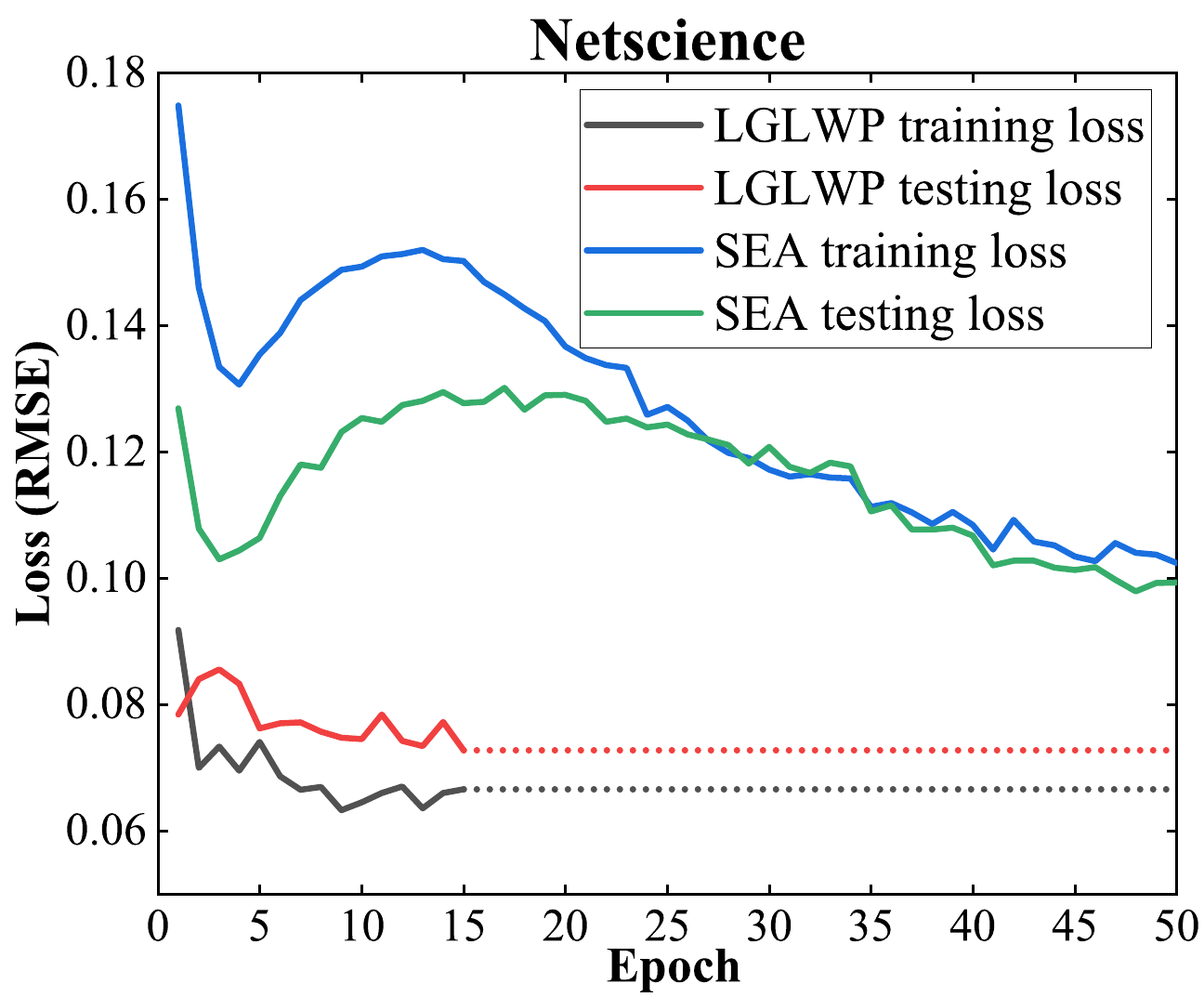}}
 	\vspace{3pt}
 	\centerline{\includegraphics[width=\textwidth]{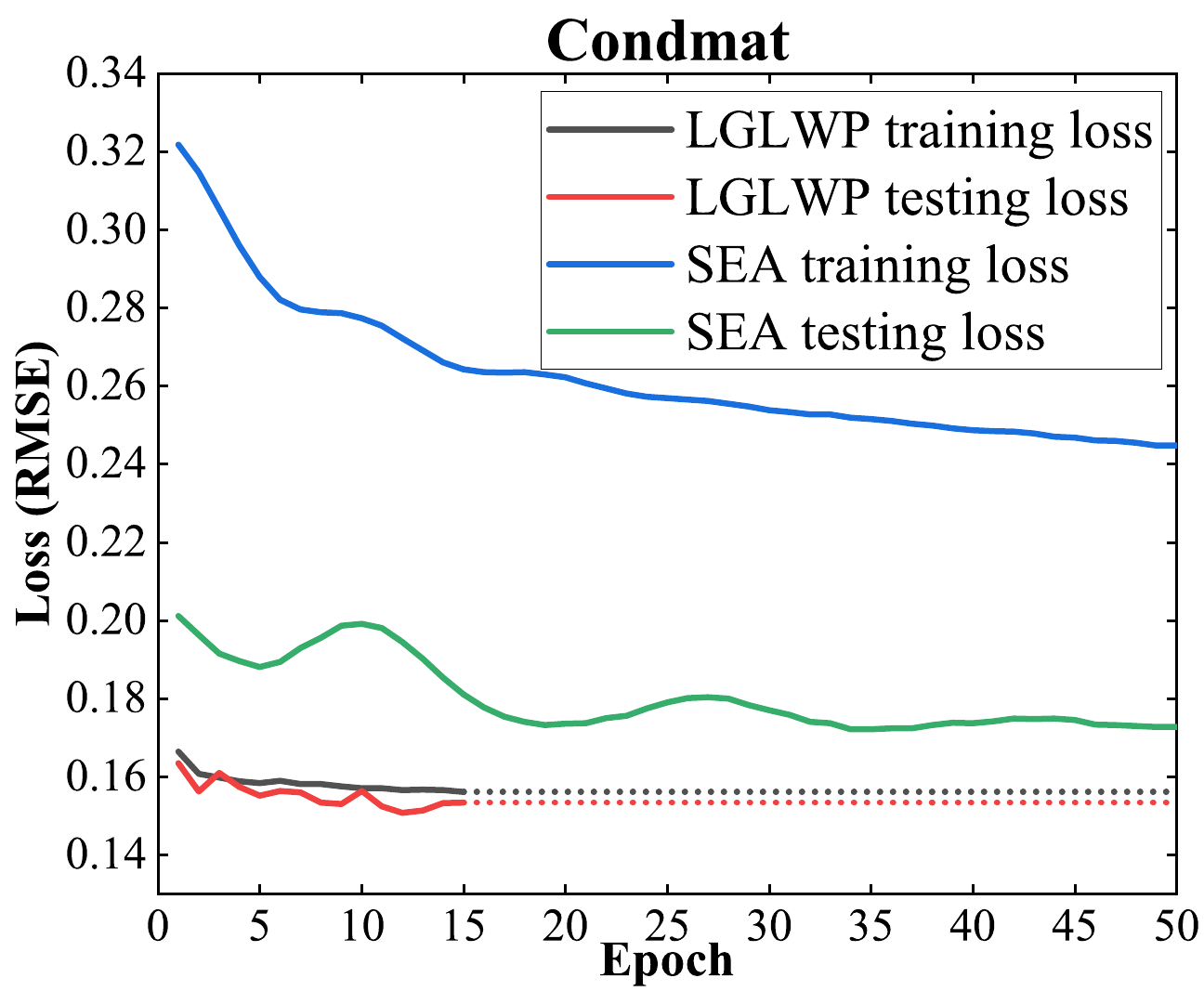}}
 \end{minipage}
	\caption{Training loss and testing loss vs. epoch for LGLWP and SEA on all the datasets.}
	\label{fig:4}
\end{figure}

To analyze the convergence speed of LGLWP and SEA, we depict the training loss and testing loss of the first 50 epochs of the two methods in Figure \ref{fig:4}. It is evident that LGLWP converges much faster than SEA across all datasets. To achieve optimal performance, LGLWP requires only about 15 epochs of training on Neural-net, C. elegans, and Netscience, and 5 epochs on P. blogs, UC-net, and Condmat. While after 50 epochs, SEA has not even converged.  To achieve optimal performance, SEA needs approximately 100 epochs on Condmat, 800 epochs on P. blogs, 500 epochs on UC-net, and 300 epochs on the remaining datasets (These results are not shown in  Figure \ref{fig:4} due to limited space). Larger networks exhibit a lower training epoch requirement for LGLWP. This is attributed to the increasing number of enclosing subgraph samples as the network size grows, facilitating our method in learning general data patterns.


\begin{figure}[htbp]
  \centering
  \subfloat
  {\includegraphics[width=0.49\linewidth]{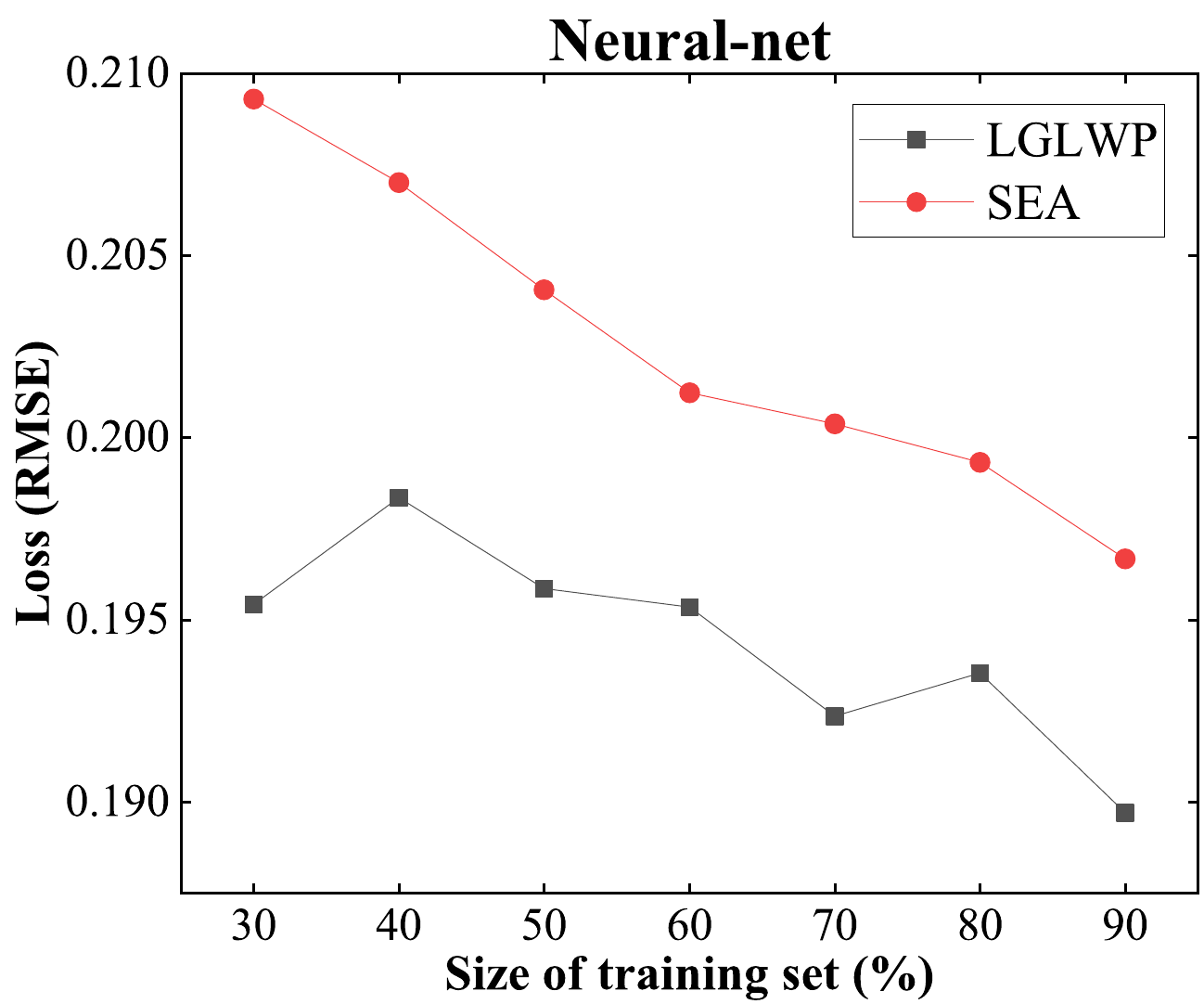}\label{fig:subfig7}}
  {\includegraphics[width=0.49\linewidth]{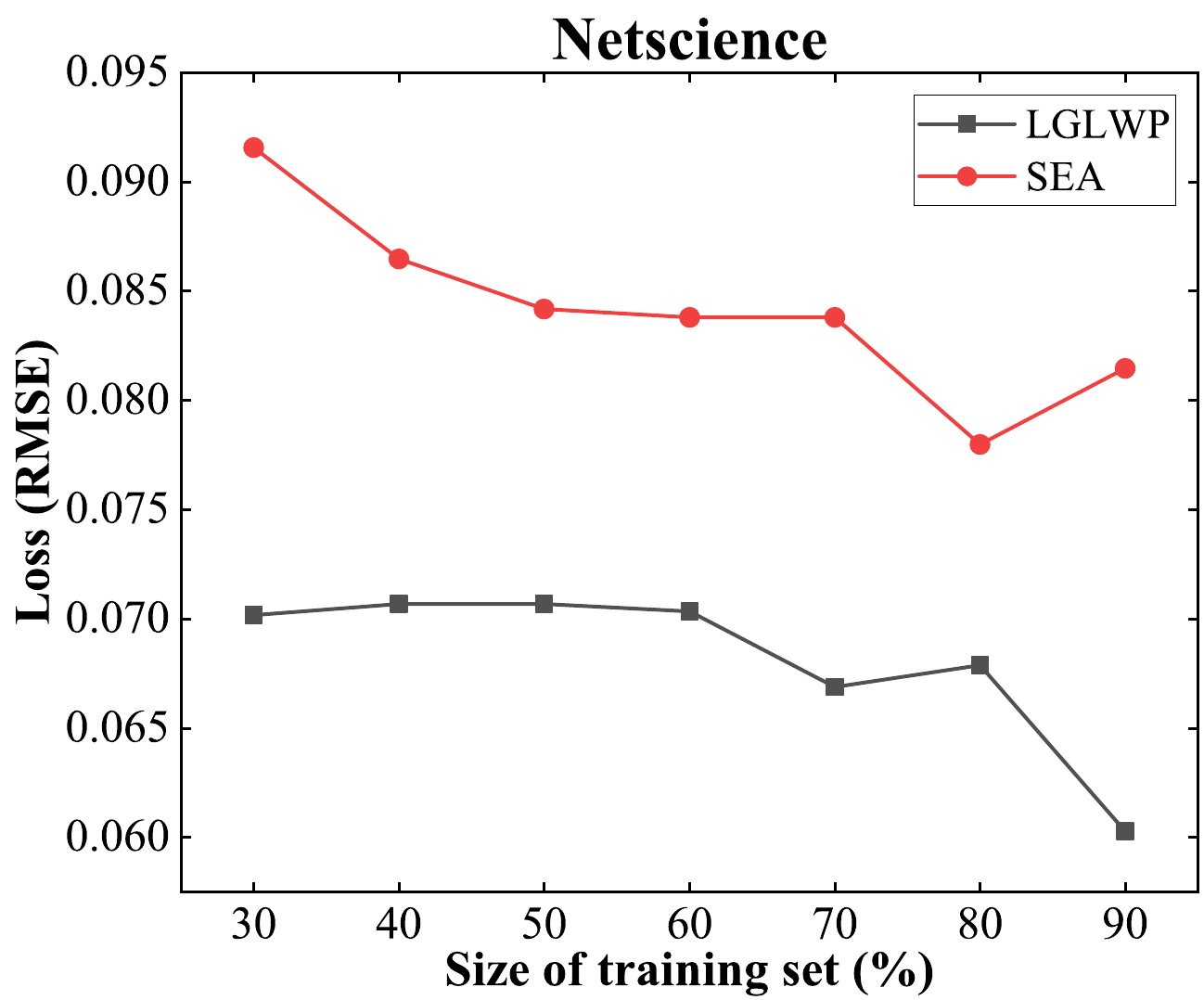}\label{fig:subfig8}}
  \quad    
  \caption{RMSE vs. size of the training set for LGLWP and SEA on  Neural-net and Netscience.  }
  \label{fig:5}
\end{figure}

To test the robustness of the prediction performance of LGLWP, we consider different sizes of the training set. The results are presented in Figure \ref{fig:5}. It is evident that LGLWP consistently outperforms SEA, as indicated by the smaller RMSE, across various training set sizes. This demonstrates the robust performance of LGLWP.

\subsection{Ablation study}
In our method, the weighted graph labeling algorithm is expected to play a big role  since it provides  a consistent way to label nodes in a subgraph according to their structural importance to link weight prediction.  To confirm the role of the labeling algorithm, we conduct ablation experiments, in which we compare this algorithm with random labeling of nodes. The experimental results on all the six datasets with the same experimental setup are given in Table \ref{tab:3}.

The results clearly demonstrate the effectiveness of the weighted graph labeling algorithm, as it achieves significantly better performance in terms of RMSE than random labeling. Due to the consideration of the balance between performance and computational cost, we only extract 1-hop subgraphs and constrain the number of subgraph nodes to be no greater than 10. It is believed that the performance of our method will increase with the expansion of subgraph size. Overall, the ablation experiments demonstrate the significance of the labeling algorithm in our method.

\subsection{Discussion}
 The experimental results validate the performance of our link weight prediction method, i.e., LGLWP. Since our method extracts only subgraphs of a limited size for prediction, the computational cost will not increase substantially with the size of the entire graph. Therefore, our method can be applied to large graphs and has even fewer training requirements in larger graphs than in smaller ones, as a larger subgraph size facilitates the learning of intricate structural patterns in the datasets. In contrast, for many other methods that need to process the entire graph, the computational cost will scale at least linearly with the size of the graph. 

In addition,  some other methods  use the graph representation learning in the prediction task.  However, they  only obtain node embeddings in the original graph and  concatenate these embeddings to represent link features, which is an indirect  way to obtain link features and may not accurately characterize the structural information of links. In contrast, our method directly maps links into low-dimensional vectors through line graph and GCN, resulting in a more effective representation of links. This makes our method more suitable for link-related prediction tasks.

\section{Conclusion}
In summary, we propose a new link weight prediction method, i.e., LGLWP, which comprises the steps of
enclosing subgraph extraction, node ordering, line graph transformation, feature learning with GCN, and regression prediction. Subgraph extraction ensures that our method relies only on local structural information to make predictions, indicating an affordable computational cost. Node ordering is employed to provide a consistent way to label nodes, reflecting the structural importance of nodes in link weight prediction. This step is critical for algorithms to learn the general patterns in the subgraphs for link weight prediction. Line graph transformation converts the link-related prediction task into a node-related prediction task, facilitating the usage of current machine learning techniques. GCN helps to learn the deep node embeddings of the line graph. Through large-scale comparative experiments, we demonstrate that our method achieves state-of-the-art prediction accuracy, which is robust regardless of the size of the training set.

\bibliographystyle{elsarticle-num}
\bibliography{my}




\end{document}